\documentstyle[prb,preprint,aps,epsf]{revtex}
\begin{document}
\draft
\title{ Non-integer flux quanta 
for a spherical superconductor}
\author{Joonhyun Yeo} 
\address{Department of Physics, Kon-kuk University,
Seoul, 143-701, Korea}
\author{M. A. Moore}
\address{Theory Group, Department of Physics,
University of Manchester,
Manchester, M13 9PL, United Kingdom.}
\date{\today}
\maketitle
\begin{abstract}
A thin film superconductor
shaped into a spherical shell at whose center lies 
the end of long thin solenoid in which there is an integer
flux $N\Phi_0$ has been 
previously extensively studied numerically as a 
model of a two-dimensional superconductor.
The emergent flux from the solenoid  
produces a radial $B$-field at
the superconducting shell and $N$ vortices in the 
superconducting film. 
We study here the effects of including
a second solenoid (carrying a flux $f$) which is inserted 
inside the first 
solenoid but passing right across the sphere.  
This Aharonov-Bohm (AB) flux does not have 
to be quantized to make the order parameter single valued.
The Ginzburg-Landau (GL) free energy  is minimized at fixed 
$N$ as a function of $f$ and it is found 
that the minimum is usually achieved  when the AB flux
$f$ is half a  flux quantum, but depending on $N$ the
minimum may be at $f=0$ or values which are not obvious 
rational fractions.
\end{abstract}
\narrowtext
A spherical shell with a magnetic monopole at its 
center   
provides a useful geometry for
numerical studies of a thin film (two-dimensional) superconductor
in a 
perpendicular magnetic field
\cite{sphere}. 
Furthermore, an investigation of the ground state of 
the vortices in this system
revealed an interesting geometric effect.
While the ground state of vortices penetrating an {\em infinite} type II 
superconductor plane 
is the well-known Abrikosov vortex lattice \cite{abrikosov},
where vortices form a triangular array, on a spherical surface
a perfect triangular lattice cannot form without the presence of
at least twelve disclination defects  and when $N$, the number
of vortices on the surface is large
other defects  appear \cite{thomson}. 

\begin{figure}
\centerline{\epsfxsize=6cm\epsfbox{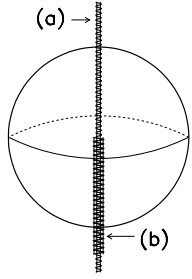}}
\vspace{10pt}
\caption{A spherical superconductor penetrated by two 
solenoids: (b) provides an approximately radial magnetic field
at the spherical surface and (a) carries an AB flux along
the z-axis.}
\label{fig1}
\end{figure}

In this paper, we descibe a situation where again  
geometry and topology play  key roles.
We consider as in
Ref.~\onlinecite{sphere} the ground state of vortices in
a spherical superconductor in the presence of a radial
magnetic field generated by a monopole
at the center of the sphere, but the vector potential describing 
the magnetic field contains also an additional
Aharonov-Bohm (AB) flux. It should be possible to realize this
system  experimentally  by inserting a solenoid into the
center of a spherical superconducting shell as the
field which emerges from the end of the 
solenoid  approximates to the field
from a magnetic monopole. One may visualize this system
as a spherical superconductor with a couple of very thin solenoids 
one of which ends at the center of the sphere and the other 
lies along the z-axis. (See Fig.~\ref{fig1}.) 

The quantum mechanical system which consists of a magnetic 
monopole {\em and} an AB flux is known to have many unusual properties
as discussed in Ref.~\onlinecite{roy} in detail. For instance, there
exist solutions to the Schr\"odinger equation for which the Dirac 
quantization of monopole charges \cite{dirac} does not hold 
even if the wavefunction is required to be single-valued.
In our system,
the monopole charges 
should be quantized, since it is determined by the 
number of vortices penetrating the superconductor as discussed below.
One of the main results of this paper is that, in the ground state, 
the system organizes
itself such that a nonvanishing AB flux is induced with its strength  
given by a fraction of the fundamental flux quantum
with the actual value related to $N$, the number
of vortices, in a way which is obscure to us. 
This 'quantization' of the AB flux differs in its origin from
other types of flux quantizations which are usually determined from
a topological consideration such 
as the requirement that the order parameter remain single valued 
as one passes round a circuit. In the present system, the flux is
quantized {\em dynamically} in the sense that it is determined by
minimizing the free energy for a superconductor.

We consider
a spherical type II superconductor of radius $R$ and width $d\ll R$
in the presence of a radial magnetic field 
${\bf H}=H(r)\hat{\bf r}$. This magnetic field is produced by 
a magnetic monopole at the center of the sphere. The system can be
described by the following Ginzburg-Landau free energy for a
(complex) superconducting order parameter $\Psi (\theta,\phi)$:
\begin{eqnarray}
F[\Psi , \Psi^* , {\bf A}] &= &d R^2 \int d\Omega\; \Large[ 
\frac{\hbar^2}{2m} |{\bf D} \Psi (\theta,\phi)|^2
+\alpha |\Psi|^2 \nonumber \\
&&~~~~~+ \frac{\beta}{2}|\Psi|^4 + \frac{1}{8\pi} | \nabla
\times {\bf A} - {\bf H} |^2 \Large] ,
\label{glfree}
\end{eqnarray}
where $d\Omega$ is the solid angle element, $\alpha,\beta$ and $m$ are
phenomenological parameters, and ${\bf D}=-i \nabla - (e^*/\hbar c)
{\bf A}$ with the vector potential ${\bf A}$ and the charge
of a Cooper pair $e^*=2e$. 
Physical properties of the system including
the effect of fluctuations are described by the partition function
given by
\begin{equation}
\label{partition}
Z=\int{\cal D}\Psi{\cal D}\Psi^* {\cal D}{\bf A}\; \exp (-F /k_B T).
\end{equation}

We study the ground state of the system by employing the mean-field 
theory in which one approximates the partition function in 
(\ref{partition}) by the saddle point values of $F$, and neglects
all the fluctuations. Furthermore we shall work in the extreme Type II
superconducting limit. 
For the vector potential (or for the magnetic induction ${\bf B}=\nabla
\times {\bf A}$), this amounts to taking the magnetic induction
such that its value is equal to the magnetic field from the 
monopole at
the superconducting surface. In terms of the vector potential,
one could take the following standard form for a magnetic monopole
with charge $g$, which has a Dirac singularity
along the negative z-axis: $A_r = A_\theta =0$,
$A_\phi=(g/r)\tan(\theta/2)$. Note that $B=g/R^2$.  
However, a more general form for ${\bf A}$
will be considered in the functional integral in (\ref{partition})
by adding an Aharonov-Bohm potential which describes the effects of an 
infinitely thin solenoid carrying flux $f$ along the z-axis
in addition to the magnetic field produced by the monopole.
On the surface of the superconductor, the vector potential 
in this case can be written as
\begin{equation}
\label{aba}
A_r=A_\theta=0,~~~A_\phi=BR\tan\frac{\theta}{2}+\frac{f}{2\pi R 
\sin\theta}.
\end{equation}
Note that except on the z-axis this vector potential gives the same 
${\bf B}$ as in the case without
the AB flux. The presence of such AB potential, however, makes a
profound change in the spectrum of the operator ${\bf D}^2$ 
in a spherical geometry \cite{roy}.

As for the order parameter,
we employ the lowest Landau level (LLL) approximation,
which one usually uses at the mean-field theory level to describe the
vortex lattice \cite{abrikosov}. We expand $\Psi$ in terms of
the eigenstates of ${\bf D}^2$ and discard all the higher eigenfunctions
except the lowest one. For a system consisting of
$N$ vortices, the flux quantization condition, which results from
the single-valuedness of the order parameter around a vortex,
gives that the total flux through the surface of the sphere,
$B(4\pi R^2) $ is equal to $N$ times the fundamental flux quantum 
$\Phi_0 =h c/e^*$.
This condition is equivalent to the Dirac's quantization condition 
for the monopole charge $g$: $2e^* g =N \hbar c$.
The normalized
LLL wavefunctions are given by \cite{roy}
\begin{equation}
\psi_m (\theta ,\phi ;a)=
h_m e^{im\phi}\sin^{|m+a|} (\frac{\theta}{2}) \cos^{|N - m -a |}
(\frac{\theta}{2}), 
\label{lll}
\end{equation}
where an integer $m$ labels the degeneracy of the LLL,
$a=-f/\Phi_0$, and
the normalization constant $h_m=[4\pi R^2 {\cal B}(|m+a|+1 ,|N-
m-a|+1)]^{-1/2}$ with the beta function ${\cal B}$.
The corresponding eigenvalues of ${\bf D}^2$ are 
$[l(l+1)-N^2 /4]/R^2$, where
\begin{equation} 
\label{eigenvalue}
l=\frac{1}{2}[|m+a|+|N-m-a|] .
\end{equation}

Since zeros of the order parameter are identified with vortices,
we might expect that $N+1$ eigenfunctions are needed
to describe the positions of 
$N$ vortices and an overall constant in the order parameter.
We note then that, because of the peculiar form
of the eigenvalues given in (\ref{eigenvalue}), one
has to include at least one eigenfunction which has a 
larger eigenvalue than the rest of the eigenfunctions.
Let us first focus on the case where $0< a <1$.
The remaining cases will be discussed later. 
We may take eigenfunctions with $m=0,1,2\cdots N$ (which is not
a unique choice as discussed below) to form
a LLL order parameter as follows:
\begin{equation}
\label{expansion}
\psi(\theta ,\phi)=P\sum^{N}_{m=0}v_m \psi_m (\theta,\phi ; a),    
\end{equation}
where $P=(k_B T hc/d\beta B e^*)^{1/4}$ and $v_m$ are complex
numbers. Inserting this into 
(\ref{glfree}) we obtain
\begin{eqnarray}
&&\frac{F[\{ v_m\} ]}{k_B T}= \alpha_T \big[ \sum^{N-1}_{m=0}|v_m|^2 
+ \epsilon (a) |v_N|^2 \big]\nonumber \\
&&+\frac{1}{2N}\sum^{N}_{k_i=0}\delta_{k_1+k_2,k_3+k_4} 
w_{1234} v^*_{k_1} v^*_{k_2} v_{k_3} v_{k_4} ,
\label{Fvm}
\end{eqnarray}
where 
\[
\alpha_T=\frac{dP^2}{k_B T} (\alpha +\frac{\hbar e^* B}
{2mc})
\]
is the dimensionless temperature which changes sign at the 
mean field transition line, $H_{c2} (T)$,  
\begin{equation}
\label{epsa}
\epsilon (a)= 1+\frac{2a}{\alpha_T}
(1+\frac{1+a}{N})\frac{dP^2}{k_B T}\frac
{\hbar e^* B}{2mc} , 
\end{equation}
and
\begin{equation}
\label{wpqrs}
w_{1234}=\frac{{\cal B}(\frac{1}{2}\sum_{i}k_i  +
2a+1,\frac{1}{2}\sum_{i}|N-k_i-a| +1)}
{\prod_{i} [{\cal B}(k_i +a +1, |N-k_i -a|+1 )]^{1/2}} 
\end{equation}
Finally, if we rescale $v_m=|\alpha_T|^{1/2} u_m$, we have
\begin{eqnarray}
&&\frac{F[\{ u_m\} ]}{k_B T}= 
\alpha^2_T \big[- \sum^{N-1}_{m=0}|u_m|^2 
- \epsilon (a) |u_N|^2 \nonumber \\
&&+\frac{1}{2N}\sum^{N}_{k_i=0}\delta_{k_1+k_2,k_3+k_4} 
w_{1234} u^*_{k_1} u^*_{k_2} u_{k_3} u_{k_4} \big],
\label{Fum}
\end{eqnarray}
where we are mainly concerned with the region
below the mean field transition
line ({\it i.e.} $\alpha_T < 0$) as the minus signs on the
right hand side indicate.
Within mean field theory the vortex lattice configuration is 
determined by minimizing the free energy (\ref{Fum})
for given $N$. We denote the minimum free energy per vortex
($F/N k_B T \alpha^2_T$) for given $a$ and $N$ by $E_N (a)$.
Once the coefficients
$\{ u_m \}$ that minimizes the free energy are obtained, the positions
of the vortices can be determined as 
elaborated later.

We first discuss an interesting symmetry relation in this system.
We note that in (\ref{expansion}) one could also use
$\psi_m$ with $m=-1,0,1\cdots N-1$ for which one eigenfunction
(with $m=-1$) has a larger eigenvalue than the rest. 
If we start from
\begin{equation}
\label{p0}
\psi (\theta,\phi)= P\sum_{m=-1}^{N-1} \tilde{v}_m
\psi_m (\theta,\phi;a) ,
\end{equation}
one can easily show that 
\begin{equation}
\label{p1}
\psi (\theta, \phi)= e^{i(N-1)\phi} P \sum_{m=0}^{N} v_m 
\psi_m (\pi-\theta ,-\phi ; 1-a ) ,
\end{equation}
where $v_m\equiv \tilde{v}_{N-m-1}$. Inserting this into 
(\ref{glfree}) we find that the free energy expression is 
just the same as (\ref{Fvm}) except that $a$ is replaced by $1-a$. 
Therefore the 
minimum free energy $E_N (a)$ as a function of $a$ has a
reflection symmetry about $a=1/2$ for $0<a<1$:
\begin{equation}
\label{sym1}
E_N (1-a)=E_N (a) .
\end{equation}
Eq.~(\ref{p1}) also suggests that the vortex  configuration
at $1-a$ can be obtained by performing the transformations; 
$\theta\rightarrow\pi-\theta, \phi\rightarrow -\phi$ on the vortex
lattice obtained at $a$.

There is also a periodicity relation with respect to $a$
in the following sense.
For $1 < a <2$ we have two choices: either $m=-1,0,1\cdots N-1$
or $m=-2,-1,0 \cdots N-2$. If we take the first and
start from (\ref{p0}), we can show that
\begin{equation}
\label{p2}
\psi (\theta, \phi)= e^{-i\phi} P \sum_{m=0}^{N} v_m \psi_m (
\theta,\phi;a-1) ,
\end{equation}
where $v_m\equiv \tilde{v}_{m-1}$ in this case. 
Proceeding as in the previous case, we conclude that
\begin{equation}
\label{sym2}
E_N (a)=E_N (a-1),
\end{equation}
and the vortex lattice has the same configuration for both cases.
One can easily see that
(\ref{sym2}) holds for all $a$ (including integral values of $a$).
Together with
(\ref{sym1}) it characterizes the
ground state energy spectrum as a function of the flux strength.
Keeping these relations in mind, we will restrict our 
discussion below only to the region $0 \leq a \leq \frac{1}{2}$.

We study the ground state of the vortex system
described by (\ref{glfree}) and (\ref{aba}) 
by numerical minimization of (\ref{Fum}).
We use a straightforward quasi-Newtonian algorithm.
The free energy (\ref{Fum}) is invariant under the transformations,
$u_m\rightarrow \exp(i\gamma )u_m $, and $u_m
\rightarrow\exp (im\gamma ) u_m$, for arbitrary real $\gamma$,
where the latter amounts to the rotation of the vortices
by an angle $\gamma$ around the z-axis. These symmetries 
enable one to set, for example, ${\rm Im} u_0 = {\rm Im} u_1 =0$ 
to reduce the number of independent variables  
from $2N+2$ ($N+1$ complex variables) to $2N$.

The first thing we notice in the numerical minimization is that
the minimum free energy occurs when $v_N =0$, {\it i.e.},
the mode that gives the larger eigenvalue than the rest 
is not present after all. This means that, for $0 \le a < 1$,
we can write the order parameter as 
\begin{eqnarray}
\psi(\theta ,\phi )&=&P \sin^a (\frac{\theta}{2}) \cos^{1-a}
(\frac{\theta}{2}) \nonumber \\
&&
\times\sum_{m=0}^{N-1}v_m h_m e^{im\phi} \sin^m (\frac{\theta}{2})
\cos^{N-1-m} (\frac{\theta}{2}) \nonumber \\
&=& C \sin^a (\frac{\theta}{2}) \cos^{1-a}
(\frac{\theta}{2}) \label{op} \\
&& \times
\prod_{k=1}^{N-1} \Big( a_k \sin (\frac{\theta}{2})e^{i\phi/2} -
b_k \cos (\frac{\theta}{2}) e^{-i\phi/2} \Big)\nonumber, 
\end{eqnarray}
where $a_k=\cos(\theta_k /2)\ \exp(-i\phi_k /2)$ and
$b_k=\sin(\theta_k /2)\ \exp(i\phi_k /2)$ with $(\theta,\phi)=
(\theta_k ,\phi_k)$ being the positions of vortices. The overall
complex constant $C$ and $N-1$ pairs of 
real numbers $ (\theta_k ,\phi_k)$
account for the original $N$ complex coefficients $v_m$. 
Note that the presence of the AB flux drives the
vortices into a peculiar configuration described by (\ref{op}),
where in addition to $N-1$ vortices at ($\theta_k ,\phi_k$)
there are {\em fractional} vortices at the north ($\theta =0$) 
and south ($\theta=\pi$) poles in the sense that the zeros are 
of order $a$ and $1-a$ respectively. In the limit where 
$a\rightarrow 0$ we obtain a configuration where one {\em full}
vortex at the south pole. As already noticed in the previous 
studies \cite{sphere} for $a=0$, one is free to fix the
position of one vortex, since the free energy
depends only on the relative positions of
vortices. Therefore the continuation from the nonzero $a$ to
$a=0$ uniquely picks the positions of vortices among the degenerate
configurations. When one continues the order parameter to 
$a=1$, one reaches the configuration where one {\em full} vortex is
on the north pole. This configuration is related to the one for
$a=0$ via (\ref{p1}), and is among the many degenerate ground states of
vortices without an AB flux. 

\begin{figure}
\centerline{\epsfxsize=7cm\epsfbox{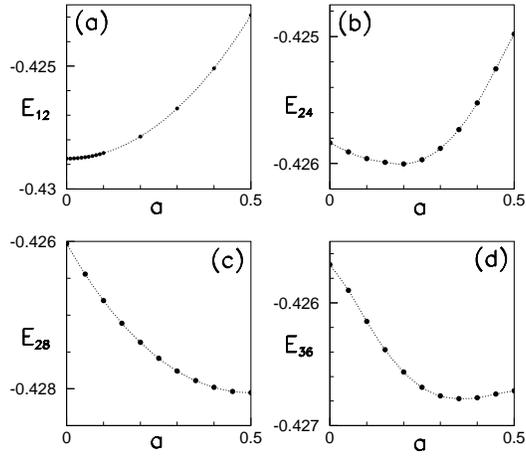}}
\vspace{10pt}
\caption{The ground state energy per vortex as a function of the
AB flux strength $a=-f/\Phi_0$ for various values
of N; (a) $N=12$, (b) $N=24$,
(c) $N=28$, (d) $N=36$.
The dotted lines are guides to the eye}
\label{fig2}
\end{figure}

\begin{figure}
\centerline{\epsfxsize=7cm\epsfbox{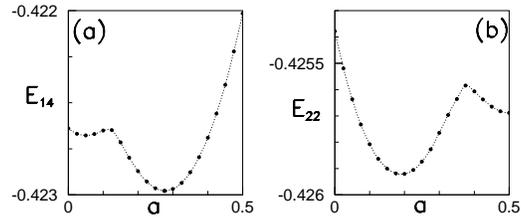}}
\vspace{10pt}
\caption{The ground state energy per vortex as a function of the 
AB flux strength for (a) $N=14$, and (b) $N=22$.}
\label{fig3}
\end{figure}

We obtain the ground state energy $E_N (a)$ as a function
of $a$ for various values of $N$. As discussed above, it is 
sufficient to consider the interval, $0 \le a \le \frac{1}{2}$. 
The main results of our
investigation is that for most of the values of $N$ we studied the 
ground state energy $E_N (a)$ has a minimum at nonzero values of $a$. 
This means 
that the vortices organize themselves such that a flux
of strength $f=-a\Phi_0$ is induced along the z-axis. 
Typical examples of energy spectra are shown in Fig.~\ref{fig2}.
For $N=1,2,\cdots 36$, we have found two exceptional cases 
at $N=14$ and $N=22$ for which the energy spectrum also shows a local
minimum or maximum as shown in Fig.~\ref{fig3}. While we have not made 
a systematic investigation of the scale of the variation of $ E_N (a)$ 
with AB flux $f$ as $N$ increases, we suspect that it is largely
$N$ independent. If so then the overall free energy change scales as
$N$, which for large N could dwarf the magnetic energy stored in 
the AB solenoid and justify our use of the word "induced``. In other
words the AB flux in this limit
could be regarded as being spontaneously generated. 
 
For some values of $N$, the ground state energy $E_N (a)$ has
a minimum at $a=0$ 
i.e at vanishing AB flux. These cases include $N=2,3,6,12,32,42 \cdots$.
We note that except for the first three, these numbers correspond to the 
so-called magic numbers found in  previous studies on the
vortex system on a sphere \cite{sphere} without the AB flux. 
It was found there that at
these numbers the ground state energy  has a lower value
than for nearby values of $N$, and the vortex
configuration in this ground state  
possesses a five-fold symmetry. 

The nonzero value of the induced AB flux in most cases is
equal to one half of the fundamental flux quantum ($a=\frac{1}{2}$).
In particular, for all odd values of $N$ up to $N=35$,
the free energy is minimized for $a=\frac{1}{2}$.
But, for some values of $N$, the value of $a$ is given by
interesting fractions other than $\frac{1}{2}$. 
These cases up to $N=36$
are tabulated in Table.~\ref{table1}. 

\begin{figure}
\centerline{\epsfxsize=6cm\epsfbox{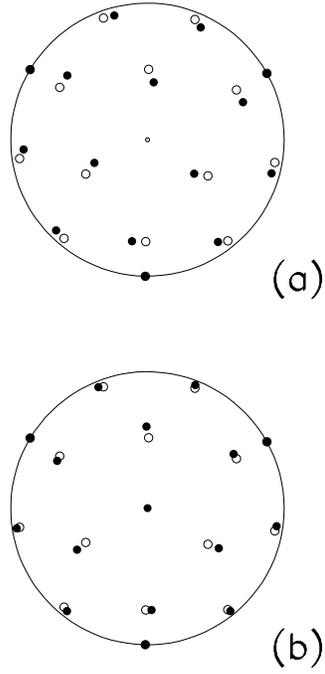}}
\vspace{10pt}
\caption{The configuration of vortices for $N=28$ and
for $a=0.05$ (filled circles) and for $a=0.5$ (open circles).
The case where $a=0.5$ has a lower energy. Figures 
(a) and (b) describe
the vortices on the northern and southern hemispheres respectively.
The poles correspond to the centers of the circles, on which 
fractional vortices are represented by the smaller size of the symbols.}
\label{fig4}
\end{figure}

The actual ground state configuration
of vortices when the AB flux is present 
is also of interest. We obtain the configuration
of vortices that minimizes the free energy 
for given value of $a$ by numerically finding the roots
$(\theta_k ,\phi_k)$ from the coefficients $\{ u_m \}$. 
We study how it changes 
when the flux strength $a$ is increased.
As remarked above, for $a\neq 0$ there are
{\em fractional} vortices at the north and south poles 
with strength $a$ and $1-a$ respectively. 
We find in general that, since
the strength of the {\em fractional}
vortex at the north pole becomes increased 
as the value of $a$ increases, 
the nearby vortices are pushed downwards. 
We expect that for most cases
one reaches a kind of balanced configuration between 
vortices on the northern and southern hemispheres 
as one approaches $a=\frac{1}{2}$,
which contributes to lowering the energy. An example of
this behavior can be seen in Fig.~\ref{fig4}. A word of
caution should be noted at this stage, however, 
that the movement of vortices as the value of
$a$ is changed is not entirely a well-defined concept, since
the vortex  configuration is only determined up to 
certain symmetry relations: A rotation around the z-axis
by any amount or a reflection in the x-z plane (which corresponds to
$u_m\rightarrow u^* _m$) does not change the energy.
In Fig.~\ref{fig4}, for instance, in order to compare the
vortex configuration for two different values of $a$,
we have rotated the vortices until the three on the equator 
in both cases coincide.     

\begin{figure}
\centerline{\epsfxsize=6cm\epsfbox{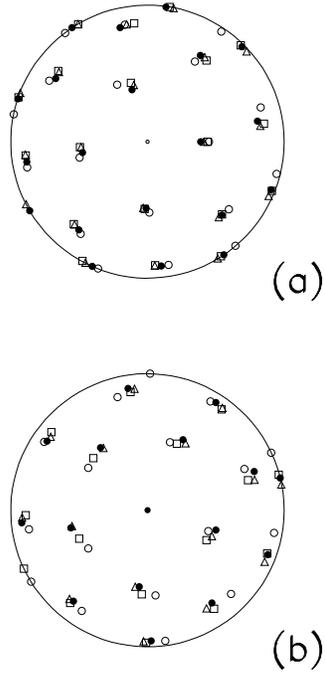}}
\vspace{10pt}
\caption{The configuration of vortices for $N=36$ and for
$a=0.15$ (filled circles), $a=0.25$ (open triangles),
$a=0.35$ (open squares), and $a=0.45$ (open circles).
The case where $a=0.35$ has the lowest energy. To compare,
we have rotated the vortices until the nearest one to the 
north pole lies on the x-axis. (a) and (b) represent the vortices on
the northern and southern hemispheres as in Fig.~\ref{fig2}.}
\label{fig5}
\end{figure}

Even if one fixes the positions vortices in an appropriate way
as in Fig.~\ref{fig4}, 
the detailed movement of vortices as the 
AB flux strength is increased is not a simple downward
translation but more complicated. This is especially true 
for the case where the minimum energy is achieved for nonzero 
fractions other than $\frac{1}{2}$ as listed in 
Table~\ref{table1}. In this case, 
while the fractional vortices on the poles give 
the nearby vortices the tendency to move downwards (or 
upwards), the other vortices also organize
themselves in a complicated way to achieve the ground state.
An example of this behavior can be seen in Fig.~\ref{fig5}. 
We believe there must be a topological reason (yet to 
be understood) which relates these nonzero fractional
values of the AB flux strength to 
the particular values of $N$.

In summary, we have presented an interesting example of an
interplay between the geometry of a superconductor and the 
vortices on it resulting in the quantization of an AB flux
with its strength given by a fraction of the
fundamental flux quantum. 
We stress again that this quantization of an AB flux has
a dynamical origin following from the minimization of the 
free energy for a vortex system. Since our analysis is based on
the mean field theory, our results will be valid only in the very 
low temperature regime. As the temperature is raised, we expect 
the quantization effect will be weakened by thermal fluctuations 
and the AB flux will fluctuate around its minimum value. 
It would also be interesting
to study possible relations of the present results to other physical 
systems exhibiting {\em fractional} vortices \cite{frac}.

We would like to acknowledge useful discussions with M.\ J.\ W.\
Dodgson. The work of one of us (J.\ Y.\ ) is supported in part
by the academic research fund of Ministry of Education, 
Republic of Korea (BSRI-97-2456).

\begin{table}
\caption{Fractional values (other than 1/2)
of the Aharobov-Bohm flux quanta 
up to $N=36$.}
\label{table1} 
\begin{tabular}{cc}
~~~~~~$N$ & $a=-f/\Phi_0$~~~ \\
\tableline
~~~~~~8 & 0.1305 $\pm$ 0.0005 ~~~\\
~~~~~~14 & 0.2775 $\pm$ 0.0005 ~~~\\
~~~~~~20 & 0.4365 $\pm$ 0.0005 ~~~\\
~~~~~~22 & 0.1895 $\pm$ 0.0005 ~~~\\
~~~~~~24 & 0.1915 $\pm$ 0.0005 ~~~\\
~~~~~~30 & 0.3250 $\pm$ 0.0005 ~~~\\
~~~~~~34 & 0.3545 $\pm$ 0.0005 ~~~\\
~~~~~~36 & 0.3615 $\pm$ 0.0005 ~~~\\
\end{tabular}
\end{table}

\newpage

\end{document}